\documentclass[prl,twocolumn,showpacs,superscriptaddress]{revtex4}

\usepackage{graphicx,bm,amssymb,amsmath}
\begin{document}

\title{Critical jamming of frictional grains in the generalized isostaticity picture}

\author{Silke Henkes}
\affiliation{Instituut-Lorentz, LION,
Leiden University, P.O. Box 9506, 2300 RA Leiden, Netherlands}
\author{Martin van Hecke}
\affiliation{Kamerlingh Onnes Laboratory, LION, Leiden University, P.O. Box 9504, 2300 RA Leiden, Netherlands}
\author{Wim van Saarloos}
\affiliation{Instituut-Lorentz, LION,
Leiden University, P.O. Box 9506, 2300 RA Leiden, Netherlands}

\begin{abstract}
While  frictionless spheres at jamming are isostatic, frictional
spheres at jamming are not.
As a result, frictional spheres near jamming do not necessarily
exhibit an excess of soft modes.
However, a generalized form of isostaticity can be introduced if
fully mobilized contacts at the Coulomb friction threshold are
considered as slipping contacts.
We show here that, in this framework, the vibrational density of
states (DOS) of frictional discs exhibits a plateau when the
generalized isostaticity line is approached.
The crossover frequency $\omega^{*}$ scales linearly with the
distance from this line.
Moreover, we show that the frictionless limit $\mu \rightarrow 0$,
which appears singular when fully mobilized contacts are treated
elastically, becomes smooth when fully mobilized contacts are
allowed to slip.
\end{abstract}

 \maketitle

The jamming transition in disordered media has been the focus of
intensive research efforts in recent years. For frictionless
spheres, the transition occurs at the isostatic point where the
contact number $z$ reaches $z^0_{\rm iso}=2d$~\cite{epitome}. This
point, known as point J,  acts as a critical point: close to it
the spectrum of vibrational modes shows an excess density of
states (DOS) at low frequencies, with a finite number of
zero-energy modes at the jamming
transition and a cross-over
frequency $\omega^{*} \sim (z-z^0_{\rm iso})$ \cite{silbert_DOS,wyart_compression}.

Friction has to be introduced to make  connection with the
granular systems studied in experiments \cite{Behringer}, in
engineering \cite{Antony_Kruyt} and in the geosciences
\cite{Marone}. For frictional spheres, the contact number at the
transition lies in the range $z^\mu_{\rm iso} = d+1 \leq z \leq
2d$, depending on the preparation method and the friction
coefficient $\mu$ of the material.
The frictional isostatic value $z^\mu_{\rm iso}$ derives from a
counting argument assuming {\em arbitrary} tangential friction
forces, and is reached in practice at jamming in the limit
$\mu\rightarrow \infty$, for gently prepared packings
\cite{kostya,Makse05,virtpaper}.

However, the magnitude of the tangential forces is limited by the
Coulomb criterion $|f_{t}|\leq \mu f_{n}$. Let $m = |f_{t}|/\mu
f_{n}$ be the mobilization of a contact, such that contacts at the
Coulomb threshold, so-called fully mobilized contacts, have $m$ $=$ $1$.
Crucially, there can be a finite fraction of contacts at the
mobilization threshold, i.e. with fixed ratio $|f_{t}|/\mu f_n$ \cite{bouchaud_leshouches,Silbert_tilt}. 
This affects the counting arguments, and suggests to consider a generalized isostaticity
condition \cite{kostya}.

\begin{figure}
 \centering
\includegraphics[width = 0.7\columnwidth,trim = 0mm 3mm 15mm 5mm, clip]{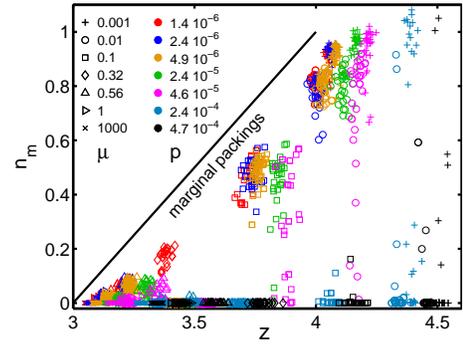}
\caption{Position of our packing configurations in the phase space
defined by $z$ and $n_{m}$ in our 2$d$ simulations. The
generalized isostaticity line $n_m = d(z-z_{\rm iso}^m)/2 = z-3 $
is shown in black. Stable packings can not exist left of this
line. Each marker denotes one of $30$ configurations at a given
$(\mu,p)$ with the color-marker combinations indicated by the
legend. For a given $\mu$ (a given symbol), configurations with
larger $p$ are to the right,
and $n_{m}$ drops to zero for the largest pressures.}
\label{fig:phasediag}
\end{figure}

We define $n_{m} \in [0, z/2]$ as the fraction of contacts per
particle at the Coulomb threshold. For frictional particles, the
tangential forces introduce $d-1$ additional force components at
each contact. Then for a packing to be stable, the $N d(d+1)/2$
rotational and translational degrees of freedom need to be
constrained by the $Nz d/2 - N n_{m}$ independent force
components. This lead Shundyak {\em et al.} \cite{kostya} to propose the \emph{generalized isostaticity} criterion
\begin{equation}
z \geq (d+1) + \frac{2 n_{m}}{d} \equiv z_{\rm iso}^{m}. \label{gen_iso}
 \end{equation}
Simulations have shown that as frictional packings are prepared
gently, the fraction of fully mobilized contacts at jamming indeed
is such that $z$ approaches the generalized isostaticity line
$z=z^m_{\rm iso}$ in the  $z$-$n_m$ plane
(Fig.~\ref{fig:phasediag}), in accord with an earlier suggestion
by Bouchaud \cite{bouchaud_leshouches}.

Fully mobilized contacts can not resist tangential perturbations,
and simulations evidence that a substantial part of the movement
of frictional piles is due to contacts failing at the mobilization
threshold~\cite{Silbert_tilt,Antony_Kruyt,Deboeuf,wyart_tilt}.
The exact values of $n_{m}$ for $0<\mu<\infty$ depend on the
preparation method~\cite{virtpaper,Deboeuf}, but in the
limit of very ``gentle'' equilibration so as to approach the
jamming threshold, they are a smooth decreasing function of $\mu$
\cite{kostya,Makse05,virtpaper}.
For 2d systems, one finds that $n_{m}$ $\to$ $1$ in the frictionless
limit   since $z \approx z^0_{\rm iso}=4$, while $n_{m}$ $\to$ $0$ in the
limit $\mu$ $\rightarrow$ $\infty$, as $z$ then approaches the
frictional isostaticity criterion $z = d+1$.

\begin{figure}
\raisebox{0.0\height}{\includegraphics[width = 0.65\columnwidth,trim = 10mm 2mm 10mm 5mm, clip]{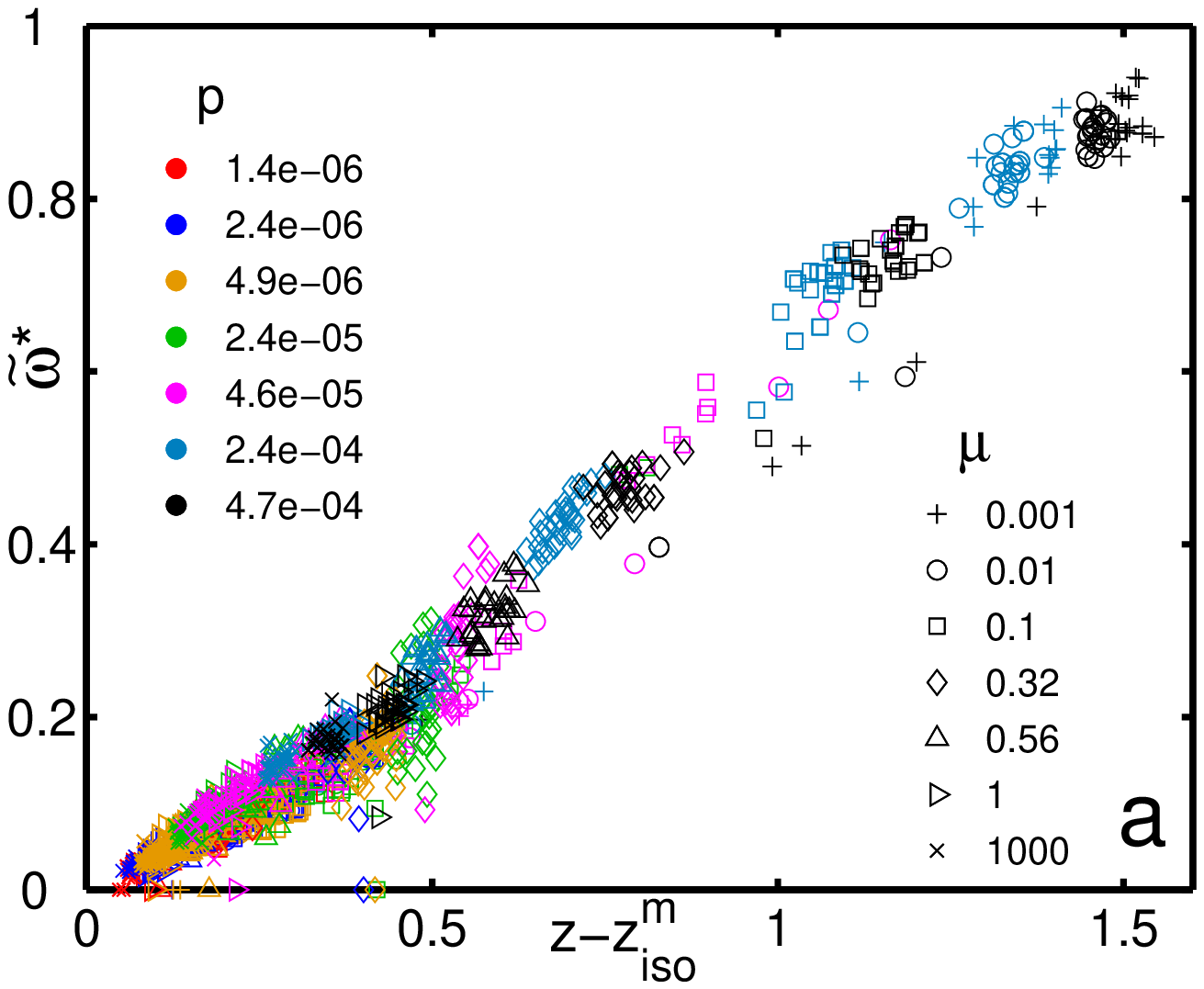}}
\raisebox{0.12\height}{\includegraphics[width = 0.264\columnwidth,trim = 96mm 147.7mm 54.5mm 33.5mm, clip]{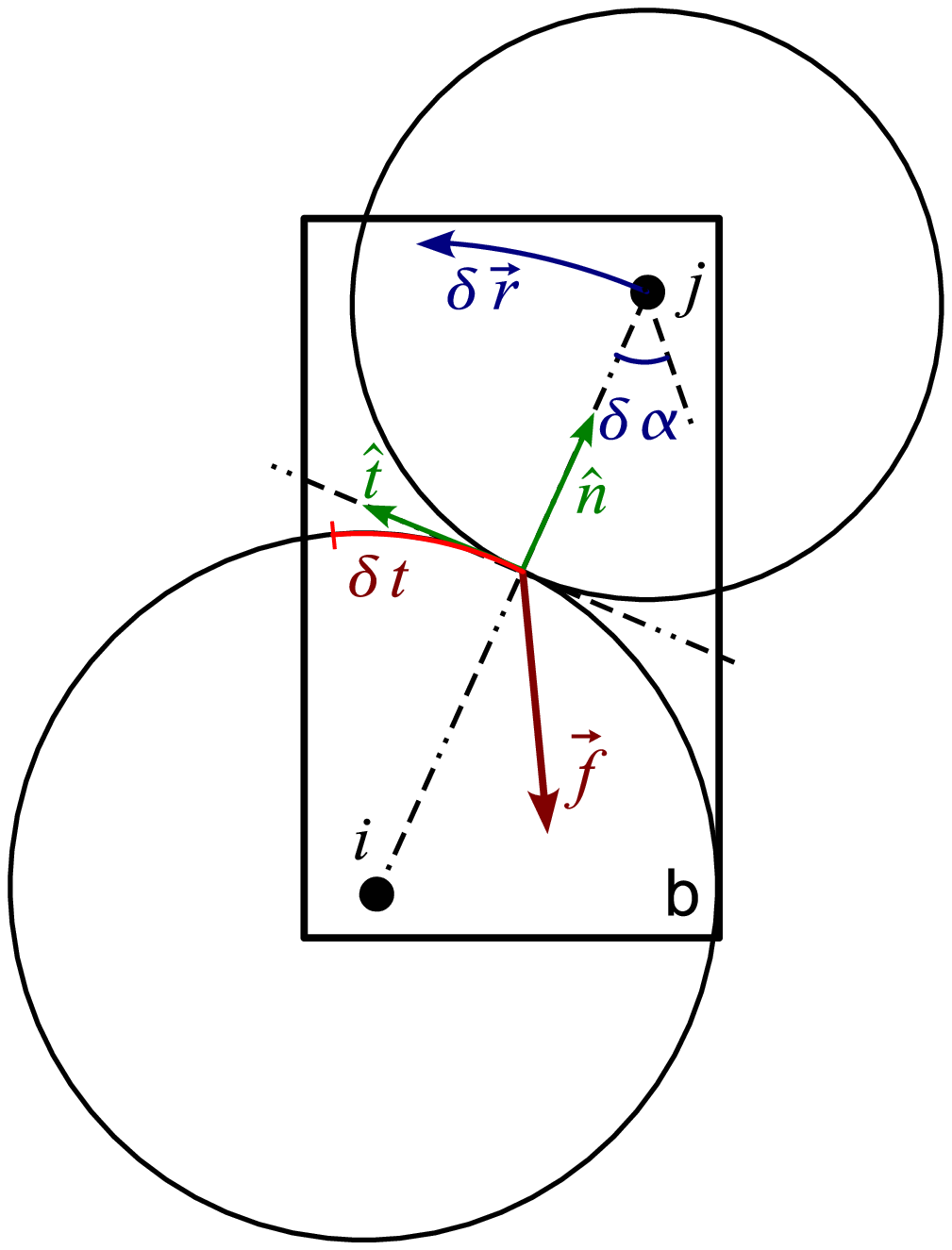}}
\caption{(a) Normalized crossover frequency $\tilde{\omega}^{*}$ where the plateau in the DOS is reached as a function of the distance $z-(3+n_{m})$ from the line of generalized isostaticity. The relation is linear, in spite of the change of the $\mu$-$n_{m}$-relation at larger pressures (see Fig.~\ref{fig:phasediag}). (b) Contact geometry illustrating the various displacements and rotations as well as the effective tangential displacement $\delta t$ of the contact point. For simplicity, we have assumed that particle $i$ is stationary.}
\label{fig:contact-wstar}
\end{figure}

In this paper, we show that the similarities between frictionless
and frictional static sphere packings, which are brought to the
foreground by this concept of generalized isostaticity, also
extend to the dynamic properties. We do so by calculating the
density of states (DOS) of frictional sphere packings, while
taking into account that contacts at the mobilization threshold
$m$ $=$ $1$ slip with unchanged tangential forces during small amplitude
vibrations. We recover the low-frequency plateau in the DOS which
is characteristic for packings of frictionless particles near
jamming. We find that the rescaled crossover frequency $\tilde{\omega}^{*}=\omega^{*}/p^{1/6}$ scales
linearly with the distance from the generalized isostaticity line,
as $\tilde{\omega}^{*} \sim (z-z_{\rm iso}^{m})$, see
Fig.~\ref{fig:contact-wstar}a. An analysis of the eigenmodes shows
that the dominant low-energy displacements consist of particles
rolling without sliding for contacts with $m$ $\neq$ $1$, and
tangential sliding at contacts with $m$ $=$ $1$.

This scenario nicely generalizes the findings for frictionless
spheres to the frictional case.
Our findings stress the crucial role of the response of fully
mobilized contacts. When instead, these are all taken to be elastic, the
inclusion of friction becomes a singular perturbation, and the
excess low-frequency modes seen in frictionless packings are
suppressed  \cite{ellak}.


{\em Simulation method} --- We perform a numerical analysis of
two-dimensional packings of frictional particles interacting with
Hertz-Mindlin forces  (during the preparation of the packings,
energy is dissipated through viscous damping  ---
see~\cite{ellak,kostya}). The Hertz-Mindlin interaction is
comprised of a normal interaction that scales as $\delta^{3/2}$,
where $\delta$ is the overlap of the particles, and an incremental
tangential interaction which is limited by the Coulomb criterion
$f_{t} \leq \mu f_{n}$~\cite{Johnson}. The packings are made at
fixed friction coefficient $\mu$ and pressure $p$.

Motivated by (\ref{gen_iso}), we map the configurations in the
phase space of Fig.~\ref{fig:phasediag} spanned by   $z$ and
$n_m$. The generalized isostaticity criterion $z=z_{\rm iso}^m$
then defines a \emph{line of marginal packings} with end points
($z=3,n_{m}=0$) and ($z=4,n_{m}=1$), so that stable packings lie
to the right of it. As in \cite{kostya}, we observe that for
sufficiently gentle preparation methods, packings approach the
generalized isostaticity line in the limit $p\rightarrow 0$.
Hence we identify this line with the (zero pressure) jamming
transition for slowly equilibrated frictional discs.
For larger pressures, packings attain higher contact numbers, and
we find a range of pressures where the distribution of $n_{m}$ is
bimodal (see e.g. the configurations for ($\mu=0.001,
p=2.4 \cdot 10^{-4}$),  due to the interplay of elastic and viscous
forces during the preparation of the packings
\cite{virtpaper}. We focus here on the behavior for $p\to
0$ at fixed viscous damping.

{\em Treatment of small amplitude vibrations
---}
The vibrational density of states is obtained by linearizing the
equations of motion around each stable state. Rattlers, which give
rise to trival zero-frequency modes, are left out from the
analysis from the start.

There are several subtleties associated with frictional response.
First of all, the nature of the ideal Coulomb condition $|f_t|
\leq \mu f_n$ implies a discontinous response for sliding
displacements at fully mobilized contacts. For displacements which lead to an increase in the tangential force $f_t$
(taking $f_n$ fixed),  the Coulomb condition then implies that the
contact slips with $f_t$ unchanged
--- this translates into a contact stiffness $k_t=0$ for this
displacement. For displacements in the opposite direction,
however, $f_t$ {\em will} decrease so that $k_{t}$ is effectively
nonzero!

This separation into two types of displacements at fully mobilized
contacts is only meaningful for static response studies. Under
vibrations, modes effectively couple through the Coulomb condition; 
moreover fully mobilized contacts dissipate energy during the
slipping half of the phase, and a decomposition into purely
nondissipative eigenmodes is not possible. In order to avoid these
complications
--- which  are artifacts of the singular nature of the ideal
Coulomb condition --- we here simply put the tangential stiffness
$k_t=0$ for all fully mobilized contacts: {\em fully mobilized
contacts are treated as slipping contacts in both directions}.
This allows us to study the DOS of these modes.

Secondly, a detailed analysis of the dynamics  shows that there always is, even at contacts which are not fully mobilized, a nonpotential term in the dynamical equations due to the tangential friction \cite{virtpaper}. The origin of this term is the change in the direction of the tangential friction force when a contact point rolls over a particle.
This effect is of the order of $f_{t}$ itself, which for our Mindlin forces scales as $\sim \delta^{3/2}$. Hence for packings close to jamming, the contribution of these terms can be neglected in comparison with the normal and tangential springs which scale as $\delta^{1/2}$; for this reason we will disregard these terms in our analysis below (similar to what is often done with the pre-stress terms in frictionless packings).

With these approximations, the equations of motion are
conservative to first order, and their structure resembles what
one finds for the vibrational properties of frictionless
particles. After we expand the equation of motion around the
equilibrium we obtain equations of motion of the form $ \delta
\ddot{r}_{\alpha} = -D_{\alpha\beta}\delta r_{\beta} +O(\delta
r^{2})$, where the    dynamical matrix  $D_{\alpha\beta}$ can,
with the present simplifications and after a rescaling of the
coordinates by the square root of the masses/moments of inertia,
be written in terms of the derivatives of an effective potential,
$D_{\alpha\beta} = (m_{\alpha}m_{\beta})^{-1/2}\frac{\partial^{2}
V}{\partial \delta r_{\alpha} \delta r_{\beta}}$
\cite{virtpaper}. For our packings with Hertz-Mindlin
forces the effective potential is given by
\begin{equation} V  = \frac{1}{2} \!\!\sum_{\langle ij \rangle} k_{n} (\vec{\delta r}.\hat{n})^{2} -  \frac{f_{n}}{r^{0}} (\vec{\delta r}.\hat{t})^{2}
+k_{t} \delta t^{2}, \label{eq:pot_energy}
 \end{equation}
where $ \delta t= (\vec{\delta r}.\hat{t}) \!-\! \left(R_{i} \delta\alpha_{i}\!\!+\!\!R_{j} \delta \alpha_{j} \right)$ is the tangential displacement of the contact point of the two particles, as illustrated in Fig.~\ref{fig:contact-wstar}(b).
Here $k_{n}$ and $k_{t}$ are the normal and tangential stiffness, respectively, which derive from the Hertz-Mindlin interaction law and which both scale as $k \sim \delta^{1/2}$ \cite{Johnson,ellak05}.
In this formulation, the vibrational spectrum is a histogram of the eigenvalues $\omega_{k}^{2}$  of $ D_{\alpha\beta} $ as a function of the associated frequencies $\omega_{k}$.


{\em Density of States} --- We have studied the vibrational
density of states (DOS) for a wide range of friction coefficients
($\mu \in [10^{-3},10^{3}]$) and pressures ($p \in
[10^{-6},10^{-3}]$). As in previous work~\cite{ellak}, we plot the
density of states as a function of the rescaled frequencies
$\tilde{\omega}=\omega/p^{1/6}$,
appropriate for Hertz-Mindlin forces which exhibit a trivial softening with 
frequencies scaling as $\sqrt{k} \sim p^{1/6}$.

\begin{figure}
 \centering
\includegraphics[width = 0.49\columnwidth,trim = 11mm 5mm 12mm 5mm, clip]{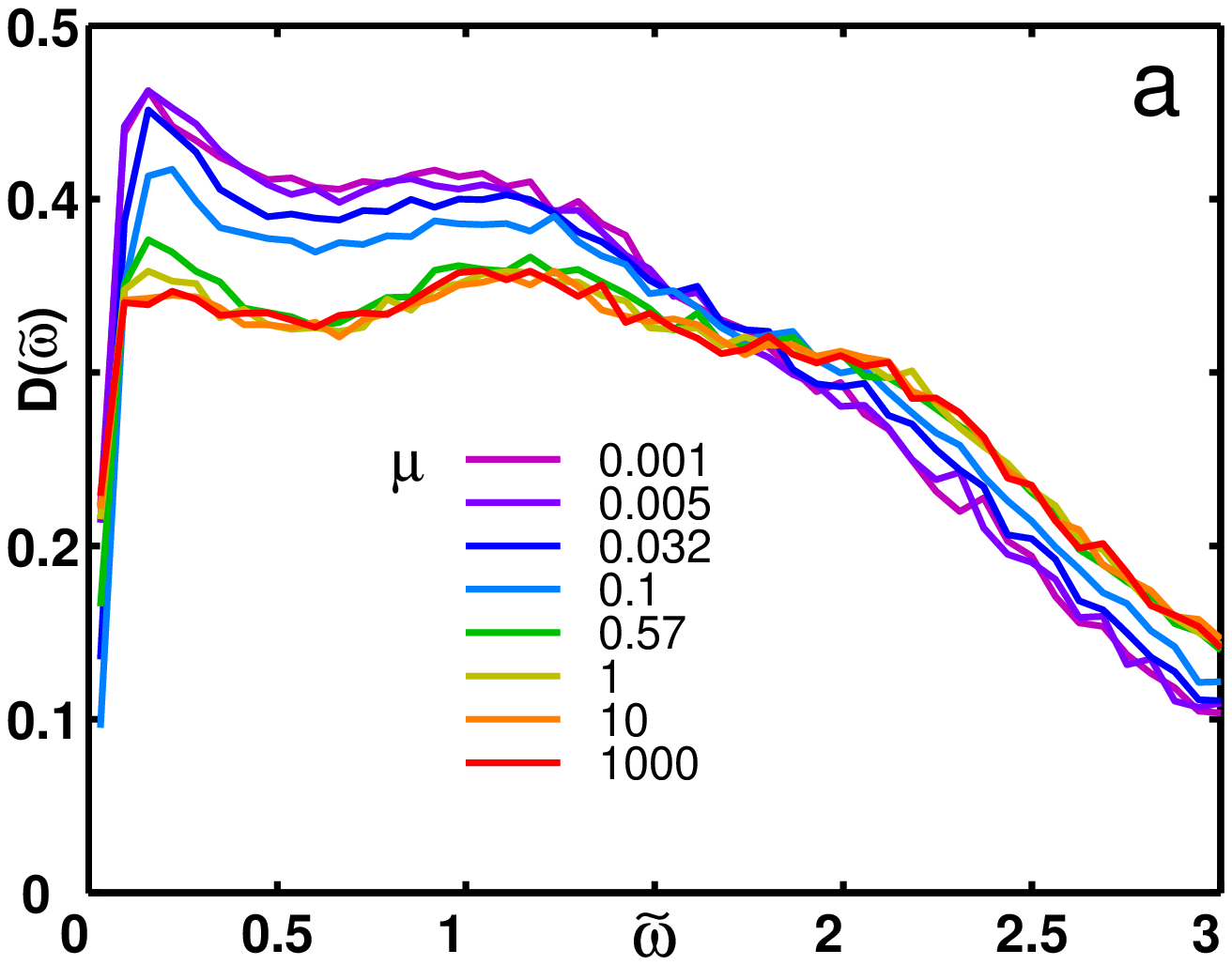}
\includegraphics[width = 0.49\columnwidth,trim = 11mm 5mm 12mm 5mm, clip]{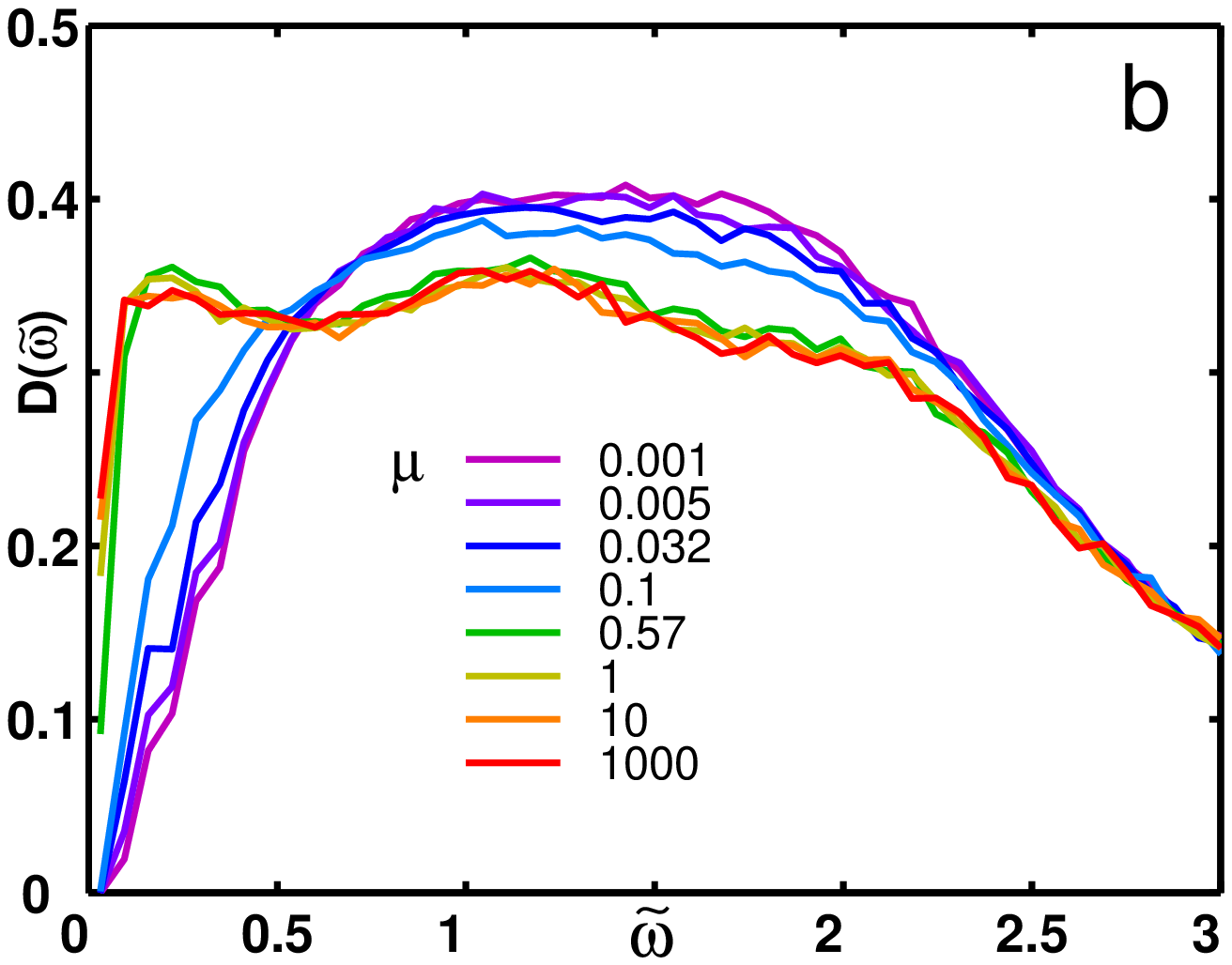}
\includegraphics[width = 0.49\columnwidth,trim = 11mm 5mm 12mm 5mm, clip]{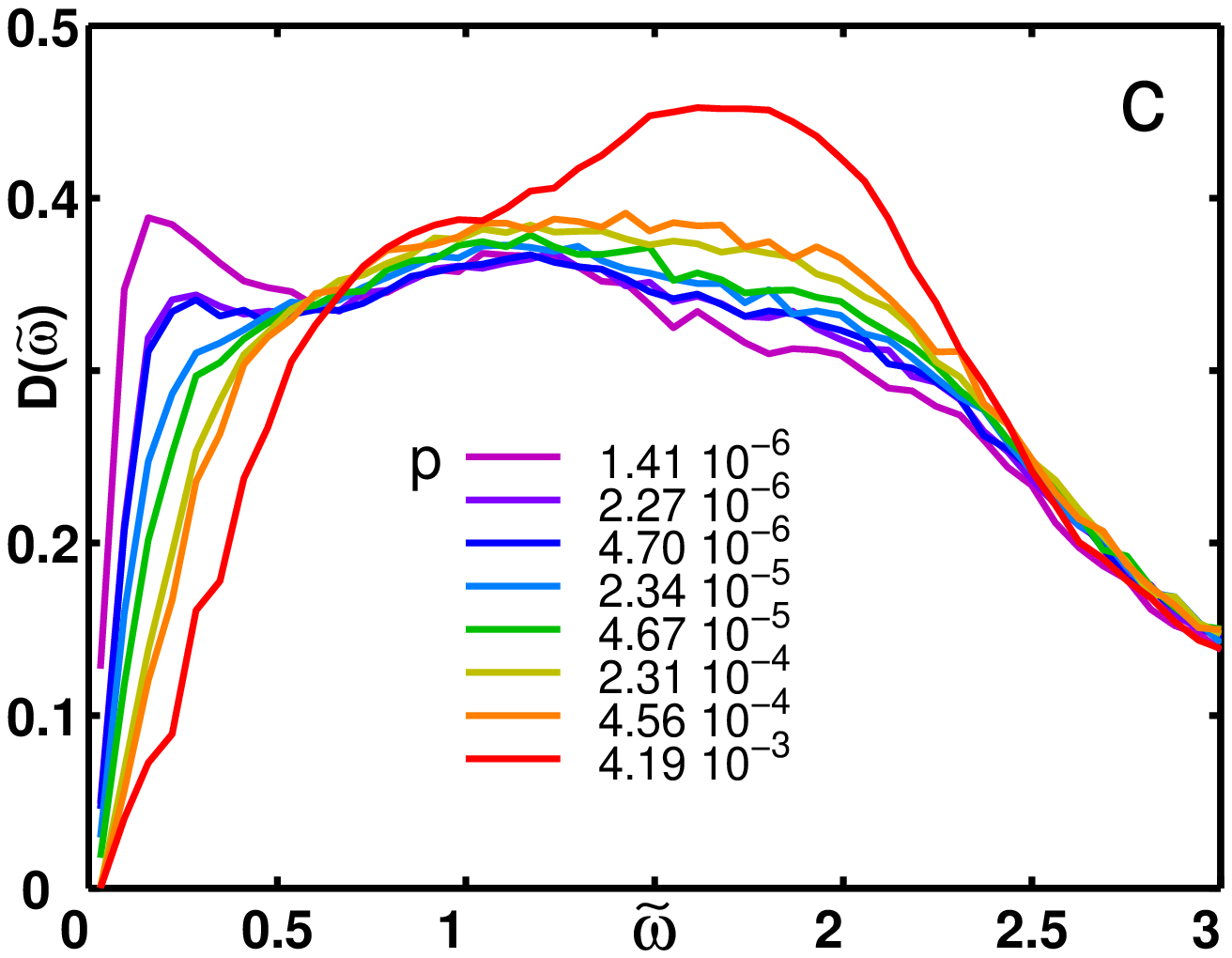}
\includegraphics[width = 0.49\columnwidth,trim = 11mm 5mm 12mm 5mm, clip]{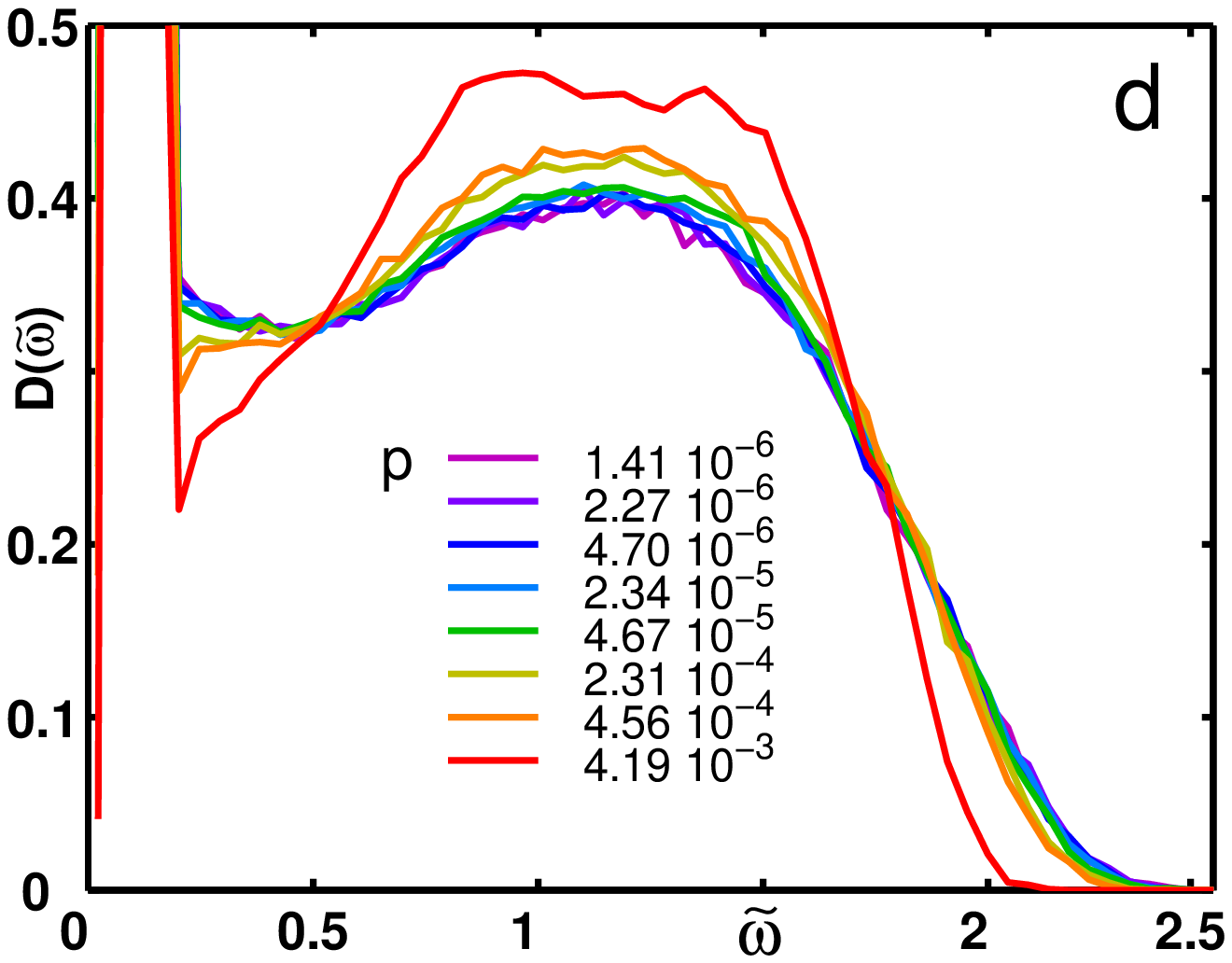}
\caption{Density of states (DOS). (a)  DOS with fully mobilized
contacts treated as slipping for the smallest pressure $p=1.41
\cdot 10^{-6}$ (approaching the line of generalized isostaticity)
for a range of $\mu$. (b) DOS for the same packings as in (a), but
with all contacts treated as non-slipping, as in ~\cite{ellak}.
(c) Illustration of the low frequency plateau developing in the
DOS as $p$ is decreased, for an intermediate friction coefficient
$\mu=0.3$ and with
$m$ $=$ $1$-contacts slipping. (d)
DOS for range of pressures and $\mu=0.001$, with the tangential
stiffness constant set to $k_{t}^{1/2} = 0.04 k_{n}^{1/2}$ (see
text).} \label{fig:DOS}
\end{figure}

Figs.~\ref{fig:DOS}a-b show the DOS for the smallest pressure
($p=1.41\cdot 10^{-6}$), i.e, close to the generalized
isostaticity line, for the full range of $\mu$. When the fully
mobilized contacts are allowed to slip, the density of states
clearly shows an excess number of low-energy modes
(Fig.~\ref{fig:DOS}a). Otherwise, when all contacts are chosen to
have a finite $k_t$,  the plateau in the DOS disappears for small
friction coefficients, where $n_{m}$ is large (Fig.~\ref{fig:DOS}b
and~\cite{ellak}).

In Fig.~\ref{fig:DOS}c, we show the evolution of the density of
states, again allowing fully mobilized contacts to slip, with
increasing pressure for $\mu=0.3$. The low-frequency part of the
DOS shows linear $D(\tilde{\omega}) \sim \tilde{\omega}$ Debye-like behavior up to
a normalized frequency $\tilde{\omega}^{*}$ which increases with pressure. In the
frictionless case, $\tilde{\omega}^{*}$ marks the frequency scale  below
which the packing can be treated like an elastic solid and above
which the DOS shows a plateau
\cite{wyart_compression}. For our frictional packings, an
extension of this argument then predicts the scaling $\tilde{\omega}^{*}
\sim z-z_{\rm iso}^{m}$ with in 2$d$ the generalized isostatic
contact number $z^m_{\rm iso}=3+n_{m}$ --- we will verify this
prediction below.

We extract $\tilde{\omega}^{*}$ in the following way: Since the DOS at
different pressures do not have similar functional forms, we
cannot rescale the integrated DOS as was done in~\cite{ellak}.
Instead, we smooth the integrated DOS to obtain an interpolated
DOS. Then $\tilde{\omega}^{*}$ is determined by the point at which
$D(\tilde{\omega})$ reaches a value of $0.2$, normalized to the height of
the plateau at $\tilde{\omega}=1$, to avoid nonlinearities in the
approach to the plateau. Fig.~\ref{fig:contact-wstar}a shows the
$\tilde{\omega}^{*}$ we obtain for different $\mu$ and $p$ as a function
of $z-z_{\rm iso}^{m}$. The relation is linear to a good
approximation, confirming our prediction.

Even if the fully mobilized contacts are allowed to slip, the
density of states for $\mu$ $\to$ $0$ is still noticeably different
from the frictionless case with $\mu=0$ (although both have an
excess of low-density modes). This is because the non-mobilized
contacts still have a finite tangential stiffness $k_t$ comparable
to the stiffness for compression of bonds $k_n$, and hence a
finite influence
--- this is even true when $p$ approaches zero and the system
approaches generalized isostaticity. Clearly, this non-smooth
behavior has its root in the singular change of the dynamical
matrix, due to the finite value of $k_t$ of many contacts
\cite{ellak}.  Fig.~\ref{fig:DOS}d illustrates that when one takes
the limit $k_{t}\to 0$ in the dynamical matrix, one recovers the
frictionless DOS, with a $\delta$-function of weight $N$ due to
trivial rotational modes at $\tilde{\omega}=0$ \cite{zz2009}.

\begin{figure}
 \centering
\includegraphics[width = 0.4\columnwidth,trim = 30mm 10mm 25mm 5mm, clip]{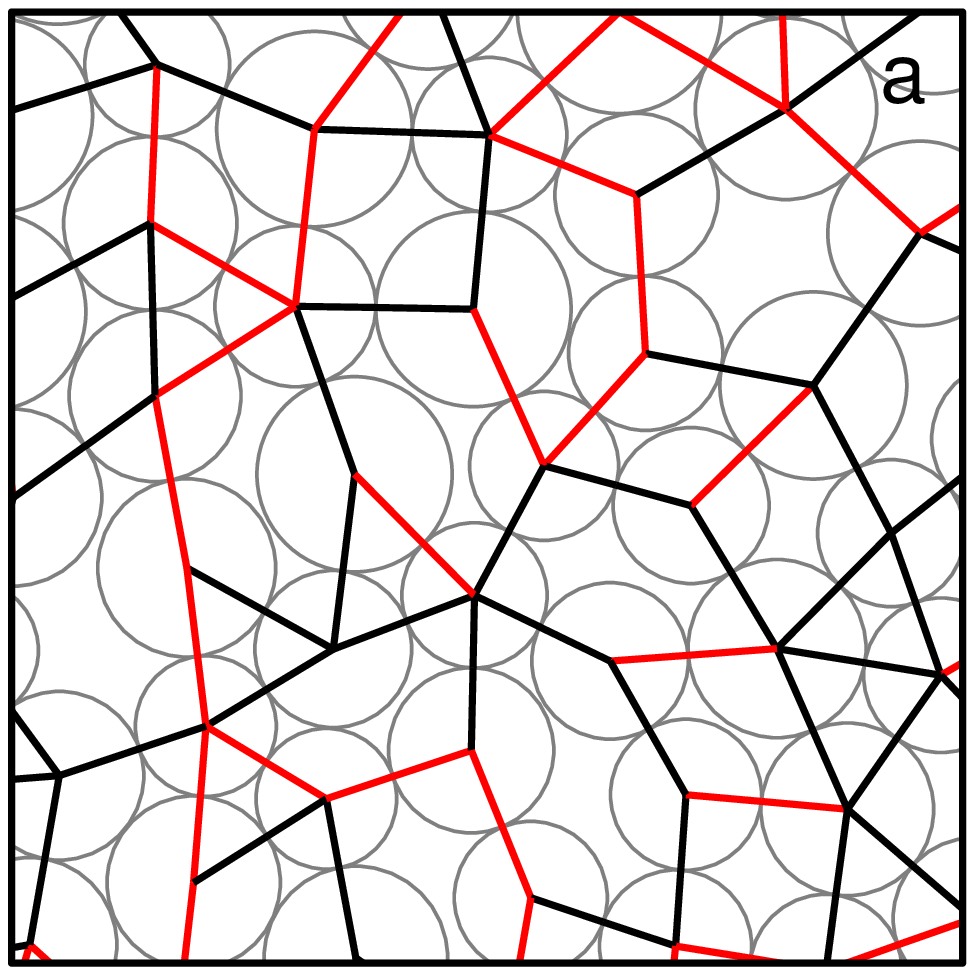}
\includegraphics[width = 0.4\columnwidth,trim = 30mm 10mm 25mm 5mm, clip]{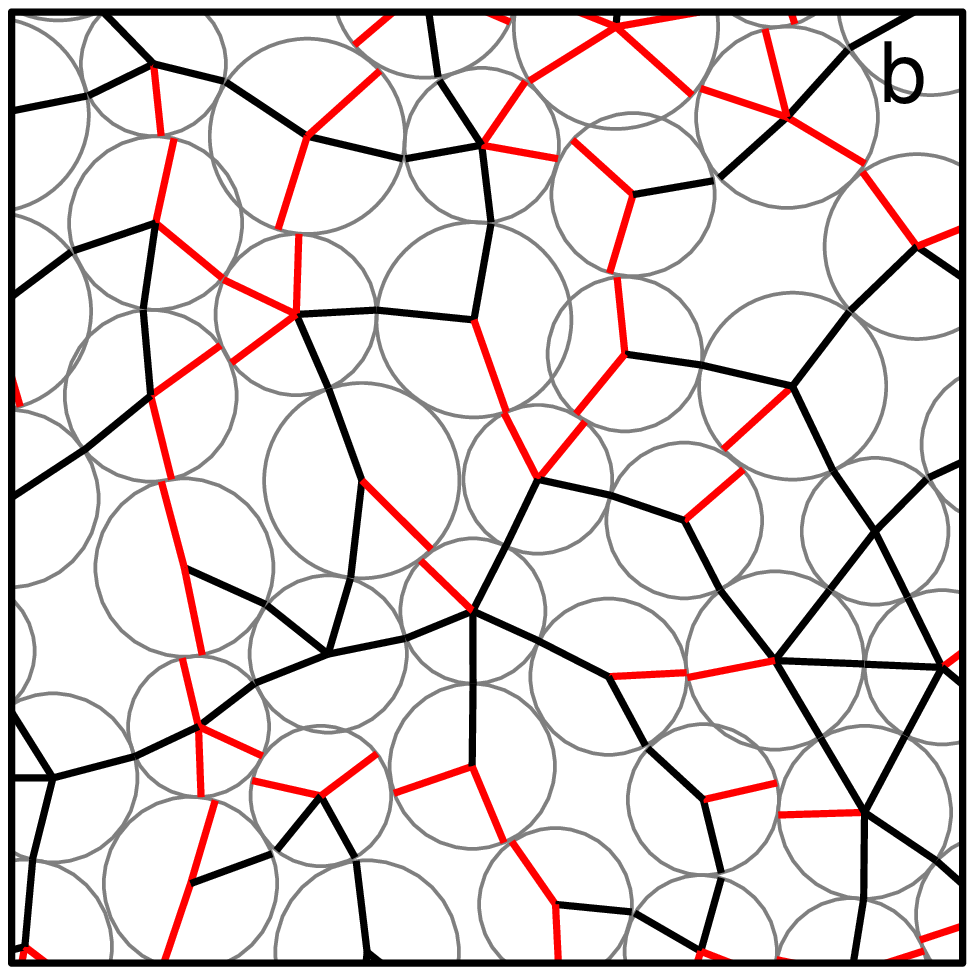}
\caption{Sample displacement for a low-energy eigenvector at
$p=1.41 \cdot 10^{-6}$ and $\mu=0.001$. (a) Contact network in the
initial state, with fully mobilized contacts in red. (b) Particles
have been displaced proportional to the mode amplitude. The lines
now link the particle centers to the position of the \emph{former}
contact points after rotation. If $\delta t \neq0$, the contact
line between particles is broken.} \label{fig:dispsample}
\end{figure}

{\em Nature of the low-energy displacements} --- We now
investigate the nature of the eigenmodes of the low-energy
eigenvalues of the dynamical matrix. Eq.~(\ref{eq:pot_energy})
predicts the nature of the lowest-energy displacements in the
limit $\tilde{\omega} \rightarrow 0$, $p \rightarrow 0$. The prefactors of
the tangential, normal and sliding displacements scale as
$\delta^{3/2}, \delta^{1/2}$ and $\delta^{1/2}$, respectively, for
a contact with $m$ $\neq$ $1$. For a $m$ $=$ $1$-contact, $k_{t}=0$.
Therefore, in the limit discussed above, the only low-energy
displacement allowed for $m$ $\neq$ $1$ is a purely tangential motion in
combination with rotations of the particles in such a way that
there is no tangential sliding. For $m$ $=$ $1$ contacts, the movement
still has to be tangential, but sliding at the contact is
permitted. A simple illustration  of the displacements in a
low-energy, low $p$ mode is shown in Fig.~\ref{fig:dispsample}:
the difference in the rotational response between slipping fully
mobilized contacts and non-slipping non-mobilized contacts is
clearly visible.

The occurrence of floppy modes at generalized isostaticity, and
their spatial structure which locally exhibits rolling and sliding
depending on the contacts mobilization, can be understood from a
counting argument. The $3N$ degrees of freedom of the packing
participating in a floppy deformation need to satisfy two groups
of constraints: setting the normal motion at each contact to zero
gives $Nz/2$ constraints, while setting the sliding motion at each
$m$ $\neq$ $1$-contact to zero gives $Nz/2 - N n_{m}$ constraints. Hence
the number of constraints of the motion is exactly equal to the
number of degrees of freedom only if $z=z_{\rm iso}^{m}=3+n_{m}$
--- floppy modes arise at generalized isostaticity. Note that the
variational argument of Wyart {\em et al.
}\cite{wyart_compression} that estimates $\tilde{\omega}^*$ in terms of
distorted floppy modes then generalizes in our case to $\tilde{\omega}^*
\sim z- z_{\rm iso}^m$.

In the limit $\mu$ $\rightarrow$ $0$, we observe that $n_{m}$ $\rightarrow$
$1$ (every other contact is fully mobilized, see
also~\cite{kostya}), so that at that point on the generalized
isostaticity line, the $ Nz/2-Nn_{m}=N$ contacts with $m$ $\neq$ $1$
equal precisely the number of rotational degrees of freedoms.
Hence there are precisely enough contacts to couple the rotations to the translations. Unlike in the case of the spherical limit of ellipsoids, where the rotational modes appear in the gap of translational modes~\cite{zz2009,mailman09}, here translations and rotations remain strongly coupled.


{\em Conclusion and outlook} --- The jamming transition of
frictional packing of spheres is more complex than the transition
at point J for frictionless spheres. However, we show that along
the \emph{generalized isostatic line} in the space spanned by the
contact number $z$ and the number of fully mobilized contacts
$n_{m}$, the system shows critical behavior similar to the
frictionless case if slipping contacts at $m$ $=$ $1$ are incorporated
into the dynamical matrix. In this case, the DOS shows a plateau
near the $z_{\rm iso}^{m}$-line, and at larger pressure, the
crossover frequency scales as $\tilde{\omega}^{*} \sim (z-z_{\rm
iso}^{m})$.

The relation between the friction coefficient and $n_{m}$ is
poorly understood except for the limiting cases $\mu$ $\rightarrow$
$\infty$ and $\mu$ $\rightarrow$ $0$. Our simulations indicate that
this relation depends significantly on the preparation method of
the sample, where packings prepared with higher viscous damping
have larger $n_{m}$ \cite{virtpaper}. The $z$-$n_{m}$
phase diagram can be a tool to understand the behavior of more
complex packings. For example, for a packing undergoing an
avalanche, the number of fully mobilized contacts increases before
the system rearranges, so that the system moves along a vertical
path in the $z$-$n_{m}$ space which eventually crosses the $z_{\rm
iso}^{m}$-line~\cite{Deboeuf}.

{\em Acknowledgement} --- We are grateful to K. Shundyak for help and use of his packings. SH gratefully acknowledges support from the physics foundation FOM.

\end{document}